\documentclass[twoside]{dis08}
\usepackage[latin1]{inputenc}
\usepackage[dvips]{graphicx,epsfig,color}
\usepackage{wrapfig,rotating}
\usepackage{amssymb,amsmath,array}

\def\lapprox{\lower .7ex\hbox{$\;\stackrel{\textstyle <}{\sim}\;$}}
\def\gapprox{\lower .7ex\hbox{$\;\stackrel{\textstyle >}{\sim}\;$}}
\newcommand{\degree}{\ensuremath{^\circ}}

\begin{document}
\title{Hadronic Final States and QCD: Summary}

%***********************************************************************
% AUTHORS INFORMATION AREA
%***********************************************************************
\author{Thomas Gehrmann$^1$, G\"unter Grindhammer$^2$,
Vivian O'Dell$^3$ and Roman Walczak$^4$
%
% Optional short acknowledgment: remove next line if non-needed
%\thanks{This is an optional funding source acknowledgment.}
%
% DO NOT MODIFY THE FOLLOWING '\vspace' ARGUMENT
\vspace{.3cm}\\
%
% Addresses and institutions (remove "1- " in case of a single institution)
1- Institut f\"ur Theoretische Physik, Universit\"at Z\"urich, 
 CH-8057 Z\"urich, Switzerland
%
% Remove the next three lines in case of a single institution
\vspace{.1cm}\\
2- Max-Planck-Institut f\"ur Physik, F\"ohringer Ring 6,
D-80805 M\"unchen, Germany
\vspace{.1cm}\\
3- Fermilab, Batavia, IL 60510-0500, USA
\vspace{.1cm}\\
4- Oxford University, Keble Road,
Oxford OX1 3RH,  UK
 \\
}
%***********************************************************************
% END OF AUTHORS INFORMATION AREA
%***********************************************************************

\maketitle

\begin{abstract}
A summary of new experimental results and
recent theoretical developments discussed
in the ``Hadronic Final States and QCD'' working group is presented.
\end{abstract}

\section{Introduction}
The study of hadronic final states at colliders provides a broad spectrum 
of insights on hadronic structure and spectroscopy and on the physics 
of quantum chromodynamics (QCD). Especially for lepton-hadron 
and hadron-hadron colliders, these results are crucial for the successful 
interpretation of any kind of 
measurement, since QCD effects are omnipresent due 
to the hadronic initial state. In this working group, a variety of new results
on many different aspects of hadronic final states were presented. These 
are summarised in this talk.

\section{Vector Boson Production}
\begin{wrapfigure}{r}{0.5\columnwidth}
\centerline{\includegraphics[width=0.43\columnwidth]{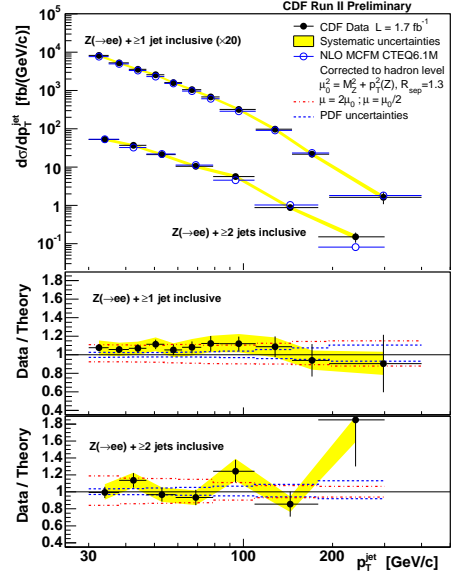}}
\caption{(a) Measured inclusive jet cross section as a function of $p_T$ in 
$Z$/$\gamma^*  \rightarrow e^+e^-$ + jets events.
Plots (b) and (c) show the data/theory ratio. Figure taken
from~\protect\cite{cooper}.}\label{fig:cooper}
\end{wrapfigure}
The measurement and modelling of inclusive single vector boson + jets production
is both a test of perturbative QCD  
and is an important background measurement
for new physics. Tevatron measurements of $Z$/$\gamma^*$ + jets and $W$ + jets
were presented~\cite{cooper}. 
Figure~\ref{fig:cooper} compares the measured inclusive 
jet cross section from events with a $Z/\gamma^* \rightarrow
e^+ e^-$ to the next-to-leading order (NLO) 
QCD prediction (open circles). The measurement for 
$N_{jet} \ge 1$ is scaled up (factor 20) for 
clarity. The shaded bands show the total systematic uncertainty except for the
5.8\% luminosity uncertainty. The
dashed and dotted lines indicate the uncertainty 
due to parton distribution functions (PDFs) and the scale variations 
of the NLO QCD predictions respectively.

The LHC will be a $W/Z$ factory, producing roughly 20 million $W$'s and 2 million $Z$'s 
visible by the general purpose detectors in each fb$^{-1}$ of data.
The ATLAS experiment at the LHC has done detailed studies~\cite{dobson} using 
both PYTHIA and ALPGEN Monte Carlo programs to extract predictions on 
reconstruction and trigger efficiency, background contributions and systematic
uncertainties for 
$W/Z$ + jets events. The dominant experimental uncertainty 
comes from the jet energy scale, and the ultimate ATLAS goal of 1\%
jet energy uncertainty would yield a systematic error of about 0.5\% on the
vector boson + jets cross sections.

For a Higgs boson mass $M_H \gapprox 2M_W$, the most promising Higgs 
discovery channel at the Tevatron and the LHC is its decay into vector 
boson pairs. In the decay into $W$-bosons, no clear mass peak 
is observable, since the neutrino from the $W$-decay leaves the detector 
unobserved. To establish a Higgs boson discovery in this channel, a 
precise understanding of the Standard Model background processes yielding 
vector boson pairs is mandatory. At present, vector boson pair production 
is described theoretically at NLO. At this order,
vector boson pairs are produced from quark-antiquark annihilation and 
quark-gluon scattering. Gluon-gluon fusion into vector boson pairs 
contributes only at next-to-next-to-leading order (NNLO), but could yield 
a potentially large contribution at the 
LHC because of the large gluon luminosity.
 
The NNLO gluon-gluon fusion contribution to vector boson pairs 
production was recently computed in~\cite{kauer}, including all vector boson 
decay information. Although gluon-gluon fusion contributes only four per cent 
to the total vector boson pair production cross section, its importance is 
substantially enhanced to above ten per cent by particle reconstruction 
cuts, and further to thirty per cent by Higgs boson search cuts.  
This observation clearly highlights the need for a full NNLO calculation 
of vector boson pair production. First steps in this direction have recently 
been completed with the calculation of the two-loop 
and one-loop squared
corrections~\cite{chachamis} to the quark-antiquark annihilation matrix 
elements in the high energy limit.

\section{Underlying Event}

\begin{wrapfigure}{r}{0.5\columnwidth}
\centerline{\includegraphics[angle=-90,width=0.45\columnwidth]{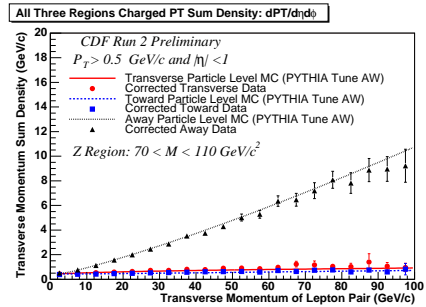}}
\caption{Underlying event study in vector boson production at CDF. Figure taken
from~\protect\cite{metha}.}\label{fig:metha}
\end{wrapfigure}
The term ``underlying event'' summarises all event activity which is 
observed besides the hard interaction, for example: multiple interactions, 
event pile-up and remnant interactions. Since this underlying activity 
results typically only in low-momentum particles, it is not accessible 
to a fully perturbative description (although at least multiple interactions 
may be calculable in the high energy limit~\cite{bartels}). At present, 
the description of the underlying event in Monte Carlo generator programs is 
based on models, such as the eikonal model of multiple
scattering~\cite{jimmy} which is used in HERWIG++~\cite{baehr1}. These models 
require tuning to experimental data, and their extrapolation from Tevatron 
to LHC energies is highly uncertain. 

In order to study and fit the underlying event in data, regions can be defined 
where the underlying event contribution is enhanced.
Figure~\ref{fig:metha} shows an example of dividing Tevatron Drell-Yan events 
into three regions with respect to the $\phi$ direction of the
lepton pair~\cite{metha}: 
the "toward" direction is defined as a cone within
$\Delta \phi < 60\degree$ to the $Z$ boson, the "away" direction is defined as a cone
within $\Delta\phi < 60\degree$ in the opposite direction, and the "transverse" direction,
 which is very sensitive to the underlying event,  is defined
as the remaining area.  The three regions are compared to PYTHIA,
 tuned to a similar 
analysis using high $p_T$ jet events (PYTHIA tune AW), and show 
good agreement.

The impact of the underlying event will be much larger at the LHC, where one
expects roughly 35 minimum bias events per bunch crossing at the design
luminosity of 10$^{34}$~cm$^{-2}$s$^{-1}$. The CMS experiment presented 
plans~\cite{bechtel}
for measuring the minimum-bias contribution using a dedicated forward
hadronic trigger and also presented studies of the remnant interactions 
besides
the hard scattering, using the same analysis strategy as at the Tevatron.
These studies are part of an overall strategy to quickly tune and understand
QCD reactions at the new energy frontier.

\section{Monte Carlo Tools}

Event generator programs are based on 
leading order matrix elements, describing a primary (low multiplicity) hard 
scattering process, which is then used as input to a parton-shower 
to generate higher multiplicities. The resulting multi-parton final state is 
then transformed into a multi-hadron final state using a hadronisation model. 
Based on Monte Carlo algorithms, these programs provide samples of unweighted 
events, which can then be further processed using detector simulation and 
event reconstruction software. They are employed very widely in all aspects 
of experimental studies at particle colliders. The programs which 
are currently used most extensively were initially developed about twenty 
years ago and have undergone continuous upgrades. Nevertheless, these 
programs now start to display serious shortcomings, since a variety of new
theoretical developments (for example, the 
matching of leading order multi-particle matrix elements onto the parton 
shower, or higher order corrections or improved 
shower prescriptions) cannot be incorporated into them. To overcome these 
limitations, several completely new Monte Carlo codes are currently being 
developed. 

The HERWIG++ project~\cite{herwig} is a new Monte Carlo generator program, 
aiming to incorporate the ideas of the well-established HERWIG generator, 
which was most widely used in the previous generation of collider
experiments. Its new fully functional release became available recently, 
and it includes, among other improvements, 
 standard interfaces to specialised matrix element generators, 
simulation of a variety of beyond-Standard-Model reactions, 
a consistent treatment of radiation off heavy particles, and the simulation 
of underlying event dynamics~\cite{baehr2}
 using an eikonal model~\cite{jimmy}.   

The SHERPA project~\cite{sherpa} is a newly developed Monte Carlo generator 
program, aiming to incorporate many of the recent new theoretical
developments: recent additions in this project include the merging of 
multi-parton matrix elements with parton showers~\cite{sherpa1}, 
a new shower model based on the dipole formalism~\cite{sherpa2},
preparations for automated NLO calculations~\cite{sherpa3}, and   
a new matrix-element generator based on improved Berends-Giele recursion 
relations~\cite{sherpa4}. 

All generic multi-purpose Monte Carlo programs are at present restricted 
to leading order in perturbative QCD. The extension of Monte Carlo programs 
to include NLO corrections is a currently ongoing activity, and has been 
accomplished for a variety of specific processes already~\cite{mcnlo}. 
For internal consistency, leading order Monte Carlo programs should 
use parton distribution functions at leading order. In fitting LO parton
distributions to deep inelastic scattering and hadron collider 
data, one observes a poor fit quality,  largely due to the high precision 
of the experimental data, which are sensitive to  higher order QCD effects
in the different observables. To improve the quality of the LO description of 
data on the proton structure, various modifications were 
suggested~\cite{sherstnev}: by easing the momentum sum rule and modifying the 
scale in the QCD evolution, it is possible to mimic some of the numerically 
dominant higher order effects. An alternative approach emerging from 
discussions in the working group would be to consider the parton distribution
functions to be an integral part of each Monte Carlo event generator, and 
to include them as  parameters to be tuned.

\section{QCD in the High Energy Limit}

In the conventional fixed-order approach to perturbative QCD, scattering 
cross sections are computed as expansion in the strong coupling constant, 
and the structure of incoming hadrons is described by parton distributions 
obtained within collinear factorisation, evolving 
according to the DGLAP evolution equations. This fixed-order approach provides 
a very successful description of a broad range of observables, it becomes 
however inappropriate if higher-order terms in the coupling constant expansion 
are enhanced by large logarithmic corrections, which can spoil the convergence
of the perturbative series. In these cases, an all-order resummation 
of the large logarithmic corrections is required to obtain reliable
predictions. In the high energy limit of QCD scattering processes, which 
corresponds to low $x$ in deep inelastic scattering, terms of the 
form $\alpha_s^n\ln^m x$ can become potentially large at all orders, thereby 
invalidating the fixed-order expansion. In this limit, collinear
factorisation, which assumes transverse-momentum ordering of initial state 
radiation, becomes equally inappropriate and should be replaced by 
transverse-momentum ($k_T$) factorisation, yielding unintegrated parton 
distributions. The resummation of large logarithms in the high energy limit 
is accomplished by the BFKL evolution equation.  

\begin{wrapfigure}{r}{0.5\columnwidth}
\centerline{\includegraphics[width=0.45\columnwidth]{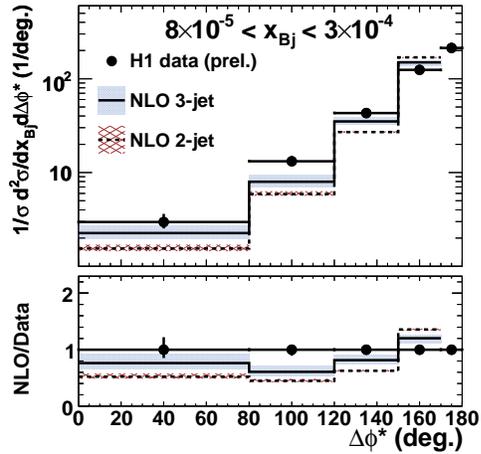}}
\caption{Dijet cross sections normalised to the visible cross section 
between $0{\degree}$ and $170{\degree}$ at HERA compared to NLO 
3-jet and NLO 2-jet calculations. Figure taken from~\protect\cite{turnau}.}\label{fig:turnau}
\end{wrapfigure}
Since the inclusive proton structure function at HERA energies is not 
sufficiently sensitive to differences between the DGLAP and BFKL approaches, 
experimental studies of BFKL resummation effects focus largely on 
specific hadronic final states at small $x$. Especially the forward jet 
cross section has turned out to be very discriminative between different
approaches. 

Data on this observable, single, double and triple differential cross 
sections are in general only poorly described by NLO QCD. Among standard
parton shower event generator programs, only ARIADNE agrees with 
observations~\cite{khein,h1-fwdjets}, after having been tuned to other
HERA data. ARIADNE is based on the colour dipole model and exhibits
BFKL-like parton showers unordered in $k_{T}$.
Differential information may also be gained from azimuthal correlations of
dijet production at low $x$~\cite{turnau}, which is not fully described
by NLO QCD as illustrated in Figure~\ref{fig:turnau}. This observable is
potentially sensitive to the unintegrated gluon distribution~\cite{hautmann},
and may enter into a global determination of unintegrated parton distribution
functions.

The BFKL equation at leading logarithmic (LL) approximation 
correctly describes the overall features of forward jet production. A fully 
reliable description can however only be attempted by including 
subleading logarithmic corrections (NLL), which are at present not yet fully 
available for jet production in deep inelastic scattering. An approximate 
NLL study~\cite{royon} does show clear improvements upon BFKL at LL, 
such as a stabilisation of the scale and scheme dependence and agreement 
with experimental data from H1 over an extended kinematical range.

To compute cross sections within $k_T$-factorisation, one needs to 
derive scattering amplitudes with off-shell initial state partons. 
To obtain those in a gauge-invariant form, it is most convenient
to couple the off-shell partons to an external current~\cite{kutak}. 
Recent calculations in this framework have focused in particular 
on final-state photon production~\cite{zotov,saleev} in proton-proton 
and photon-proton collisions. Since these calculations do not provide 
 the full 
final state information on all partons, issues like photon isolation and 
infrared cut-offs are still controversial.

\begin{wrapfigure}{r}{0.5\columnwidth}
\centerline{\includegraphics[width=0.45\columnwidth]{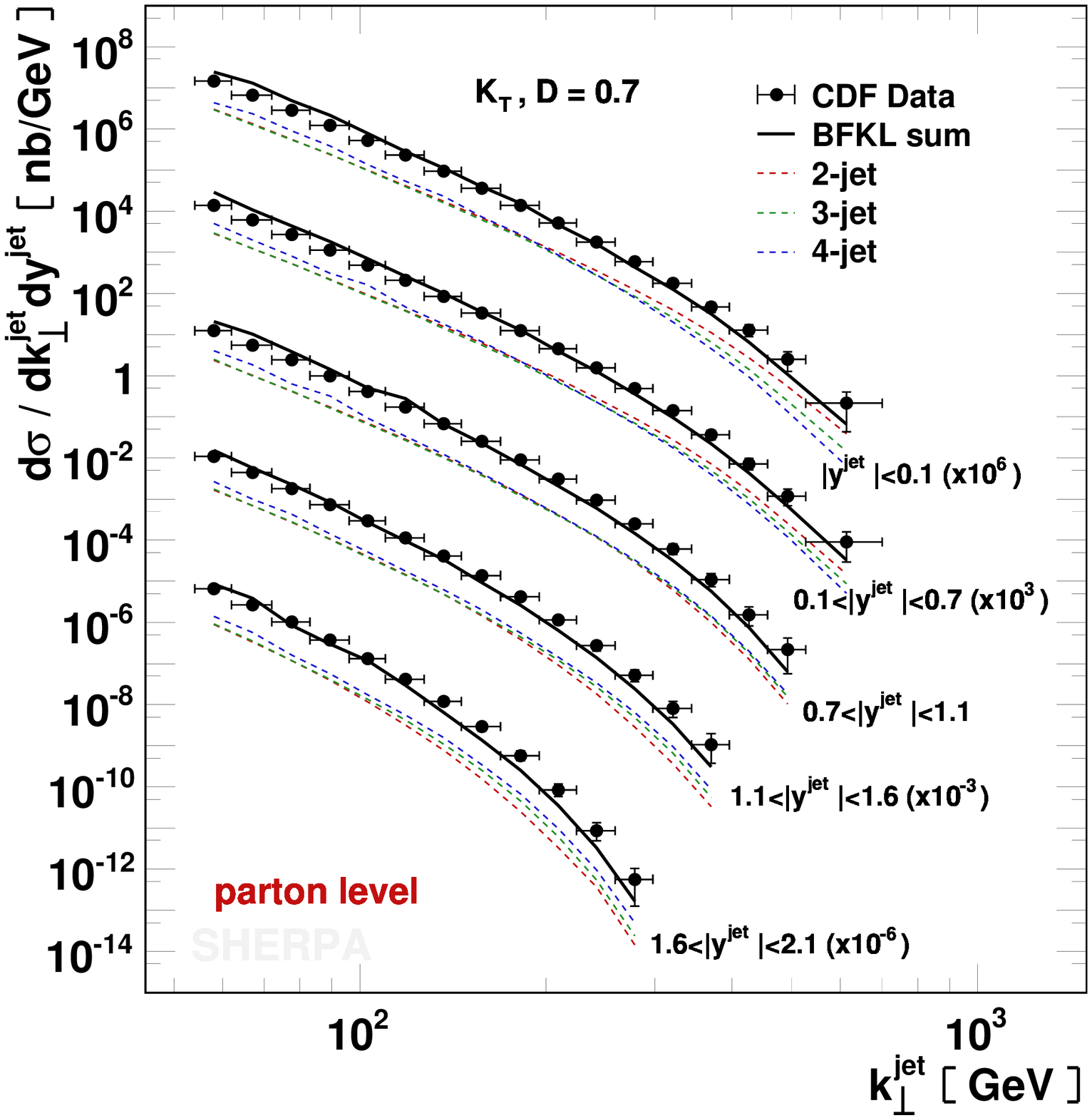}}
\caption{Inclusive jet cross sections at Tevatron compared 
to LL BFKL parton shower predictions. Figure taken
from~\protect\cite{hoeche}.}\label{fig:hoeche}
\end{wrapfigure}
The BFKL evolution equation at leading logarithmic accuracy can be 
reformulated as a parton shower by interpreting its evolution kernel 
as splitting probability~\cite{hoeche}. In this formulation, constraints from 
energy-momentum conservation are implemented in a straightforward manner.
The unintegrated parton distribution functions required as initial 
conditions are inferred from ordinary DGLAP parton distributions 
by undoing the last branching in the DGLAP evolution. With this parton 
shower formulation, leading logarithmic BFKL predictions can be made for 
a great variety of observables. For example, inclusive jet production at the 
Tevatron is correctly described over a wide kinematical range, 
 Figure~\ref{fig:hoeche}. It can be clearly seen that in certain kinematical 
regions  different jet multiplicities can be of comparable 
numerical magnitude in the BFKL approach. 

The BFKL formalism can also be applied directly to
approximate the matrix elements for hard scattering needed
for the prediction of multi-jet events. When supplemented by
simple constraints on the analytic behaviour of the
amplitudes~\cite{white}, a good approximation of the corresponding full
leading order matrix
elements is acheived, while allowing for an all-order resummation of the
hard perturbative corrections.

A complementary approach to the high-energy limit of QCD is the eikonal 
approximation, which allows to compute soft-gluon corrections to all 
orders in the coupling constant. This approximation is 
applied for example in pair production~\cite{kidonakis} of heavy quarks.

\begin{wrapfigure}{r}{0.5\columnwidth}
\centerline{\includegraphics[width=0.45\columnwidth]{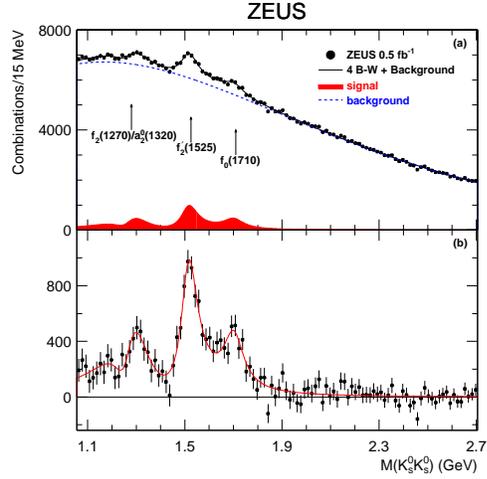}}
\caption{The measured $K^{0}_{S}K^{0}_{S}$ invariant-mass distribution. 
Figure taken
from~\protect\cite{karshon}.}\label{karshon_fig}
\end{wrapfigure}
\section{Hadron Spectroscopy}
Hadron spectroscopy has been playing an important role in understanding strong 
interactions. Recently, ZEUS has studied $K^{0}_{S}K^{0}_{S}$ final states
in $ep$ collisions at HERA \cite{karshon}. Three enhancements in invariant-mass distribution 
were observed, as shown in Figure~\ref{karshon_fig}. 
They correspond to $f_{2}(1270)/a^{0}_{2}(1320)$, 
$f^{'}_{2}(1525)$ and $f_{0}(1710)$ mesons. The interference pattern, 
predicted by SU(3) symmetry, was taken into account fitting the invariant-mass distribution. 
The $f_{0}(1710)$ state, which has a mass consistent with a $\rm J^{PC} = 0^{++}$ 
glueball candidate, is observed with 5 standard-deviation statistical 
significance. However,  if this state is the same as that seen in 
$\gamma \gamma \rightarrow K^{0}_{S}K^{0}_{S}$ \cite{TASSO_L3}
it is unlikely 
to be a pure glueball state.

A broad spectrum of results on hadronic interactions at 
low energies, including nuclear 
form factors and resonances~\cite{devita} and colour 
transparency~\cite{gilfoyle}  is currently being obtained at TJNAF. 

\section{Heavy-Ion Collisions}

\begin{wrapfigure}{r}{0.5\columnwidth}
\centerline{\includegraphics[width=0.45\columnwidth]{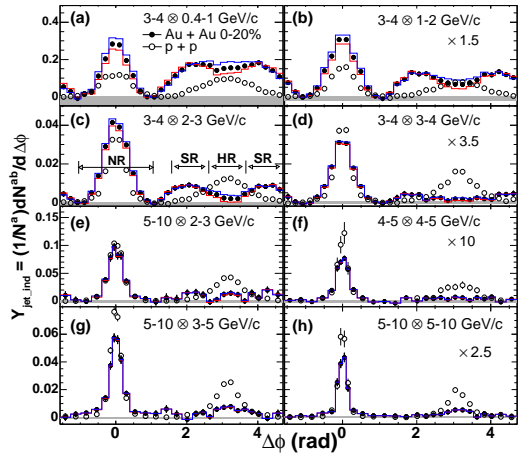}}
\caption{Per-trigger yield versus $\Delta\phi$ for 
various trigger and partner $p_T$ arranged by increasing pair proxy 
energy. Figure taken
from~\protect\cite{caines}.}\label{fig:caines}
\end{wrapfigure}
At the Relativistic Heavy-Ion Collider, jets have been used to study
the high energy density matter created in nuclear collisions. Defining
regions in a similar way as in underlying event studies in $p\bar{p}$ collisions,
one can study the jet evolution in the quark matter. Figure~\cite{caines}
compares the jet evolution in $pp$ and Au-Au collisions for increasing
jet energy (here a jet is defined as a high energy hadron). In this figure, 
solid histograms (shaded bands) indicate elliptic flow
model uncertainties. Arrows in (c) show the ``head" region (HR) the
``shoulder" region (SR) and the ``near-side" region (NR). 
 The $pp$ 
data shows a typical two peak structure due to  back to back dijets. The Au-Au
spectra reveal jet quenching in the medium as the $p_T$ increases, 
and in addition, they show a  more complicated jet structure evolution with 
$p_T$, with prominent peaks in the ``shoulder" regions at fixed positions
-- a feature expected from a medium-induced Mach shock.

\section{Hadron Fragmentation}

\begin{wrapfigure}{r}{0.5\columnwidth}
\centerline{\includegraphics[width=0.45\columnwidth]{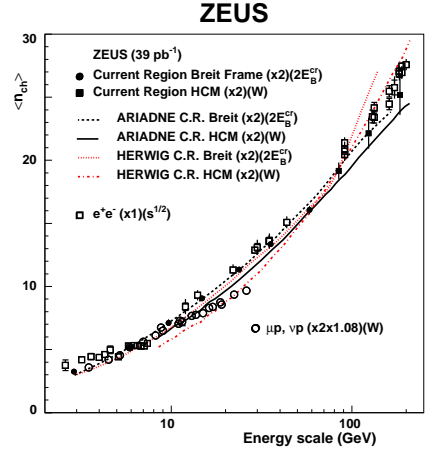}}
\caption{Mean charged multiplicity. 
Figure taken
from~\protect\cite{ZEUSchp}.}\label{Tymieniecka_fig}
\end{wrapfigure}
Charged particle production in DIS at HERA was studied in terms 
of multiplicity and  
scaled momentum distributions \cite{ZEUSchp,H1chp}. 
When an appropriate energy scale is used, $ep$ data can be consistently 
compared with data from $e^{+}e^{-}$, $\mu p$ and $\nu p$ 
scattering over a wide energy range. In all cases, similar behaviour is observed, 
supporting quark fragmentation universality, see Figure~\ref{Tymieniecka_fig}.

\begin{wrapfigure}{r}{0.5\columnwidth}
\centerline{\includegraphics[width=0.45\columnwidth]{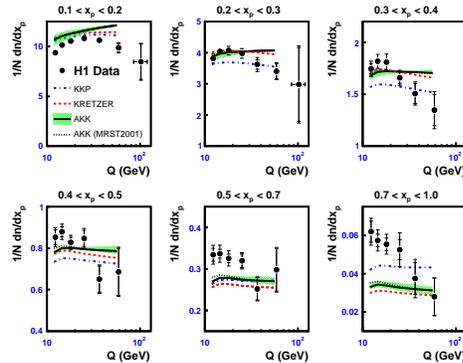}}
\caption{Normalised distributions of the scaled momentum. 
Figure taken
from~\protect\cite{H1chp}.}\label{Traynor_fig}
\end{wrapfigure}

Models, implemented in LO matrix element Monte Carlo programs, describe 
data reasonably well but 
NLO QCD calculations,
using three different fragmentation functions, fail to describe the scaling violations seen
in the data \cite{H1chp}, see Figure~\ref{Traynor_fig}.

The production of $K_{S}^{0}, \Lambda$ and $\bar{\Lambda}$  
in DIS at HERA has been investigated by H1 \cite{falkiewicz}. 
The model predictions, based on leading order Monte Carlo programs 
in general are able to describe the overall features of the measurements, 
however they fail in details, in particular in describing $\Lambda$ and $\bar{\Lambda}$ 
production in the current region of the Breit frame. It was also found that a
constant strangeness fraction in hadron fragmentation fails to fit all the data. 

Data from $pp$, $e^{+}e^{-}$ and heavy ion scattering has been used 
to test the limiting fragmentation hypothesis \cite{B-Jmodel}, and a prediction has been made for 
$ep$ DIS \cite{tymieniecka}. 

Bose-Einstein correlations of hadron pairs at HERA have been studied 
by HERMES \cite{gapienko} for nuclear targets ranging from hydrogen to xenon. 
It was found that the parameters describing the correlations neither depend 
on the nuclear target nor on the hadronic invariant-mass. 

Momentum distributions of identified hadrons  
can be used to extract hadron fragmentation functions, which describe 
the differential probabilities for specific 
parton-to-hadron transitions. These fragmentation functions obey 
DGLAP evolution equations with timelike splitting functions. At present,
these timelike splitting functions are known to NLO,  and 
calculations to NNLO are in progress~\cite{moch}. 
Using extensive data sets taken in 
$e^+e^-$,  $pp$  and $p\bar{p}$
collisions, and including hadron mass corrections, 
a new global extraction of 
parton fragmentation functions to $\pi^\pm$, $K^\pm$, $p/\overline{p}$,
$K_S^0$ and $\Lambda/\overline{\Lambda}$ was performed 
recently~\cite{kniehl}. Compared to earlier studies, especially the 
determination of non-singlet quark fragmentation functions is improved 
by charge asymmetry $pp$ data from RHIC. 
Unfortunately, data on momentum distributions of 
identified light hadrons from HERA have up to now 
only been released by the fixed-target HERMES experiment, while 
H1 and ZEUS measurements of charged and neutral light hadron 
production released up to now were not suitable for the extraction of 
fragmentation functions.

\begin{wrapfigure}{r}{0.5\columnwidth}
\centerline{\includegraphics[width=0.45\columnwidth]{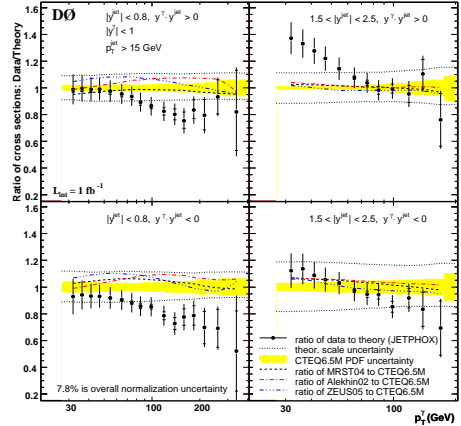}}
\caption{Ratio of measured and predicted 
cross section for $p\bar{p}\rightarrow\gamma + jet + X$ as
a function of $p_T^\gamma$. Figure taken
from~\protect\cite{campanelli}.\label{fig:campanelli}}
\end{wrapfigure}
\section{Isolated Photons}
Direct photons coming from the partonic hard scattering are
 a powerful probe of 
of the dynamics of hard QCD interactions. At Tevatron energies, the 
direct photon + jet cross section measurement is 
sensitive to the gluon density and thus can
be used to complement the HERA data quark on the gluon at low $x$ by 
constraining
the gluon contribution
at large $x$. In addition, photon final states are predicted in
new physics models such as SUSY, Extra Dimensions, Technicolour, etc.
The DZero experiment has measured~\cite{campanelli} 
the isolated (hence enhanced direct)
$\gamma$ + jet inclusive cross section in four regions
 of jet and photon rapidity
and compared them with theory using JETPHOX~\cite{jetphox} and CTEQ6.5M PDFs
as shown in Figure~\ref{fig:campanelli}.  
The two dotted lines show the effect of theoretical scale variations by
a factor of two; the shaded region indicates the CTEQ6.5M PDF uncertainty, 
and the  dashed and
dashed-dotted lines show ratios of the JETPHOX predictions with MRST 2004, 
Alekhin and ZEUS 2005 PDF sets to CTEQ6.5M. The systematic uncertainties have 
large bin-to-bin correlations in $p_T^{\gamma}$. 
An additional 7.8\% normalisation 
uncertainty  on the data points is not shown.

\begin{wrapfigure}{r}{0.5\columnwidth}
\centerline{\includegraphics[width=0.45\columnwidth]{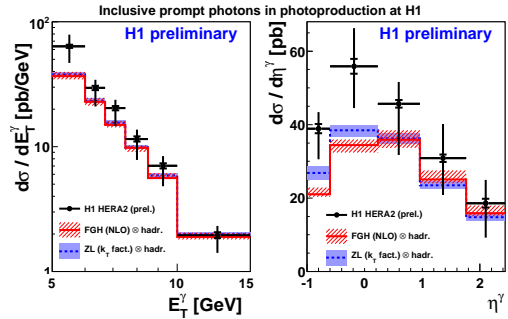}}
\caption{Inclusive prompt photons in photoproduction at HERA. Figure taken
from~\protect\cite{nowak}.}\label{fig:nowak}
\end{wrapfigure}
While there is  overall agreement with
the theoretical prediction, and  the theory successfully models the
$p_T$ distributions for the isolated $\gamma$+jet measurements at HERA, 
the theory does not describe well the 
$p_{T}^{\gamma}$ dependence of the cross sections as measured in $p \bar{p}$
at the Tevatron. Neither reasonable variations in fragmentation functions
nor contributions from threshold resummation are able to improve the
data description from theory. Clearly this is an important measurement to 
be understood in preparation for data taking at LHC.

First results on prompt photons in photoproduction by the H1
collaboration making use of the full HERA-II statistics
 were presented~\cite{nowak}. They cover the phase
space $5 < E_{T}^{\gamma} < 15$~GeV and $-1.0 < \eta^{\gamma} < 2.4$ in the laboratory frame. The isolation criterion for the prompt photon is defined by
the requirement that $E_{T}^{\gamma} / E_{T}^{\gamma -jet} > 0.9$. The data 
are above the two QCD predictions, particularly at low $E_{T}^{\gamma}$ as 
can be seen in Figure~\ref{fig:nowak}. The FGH prediction~\cite{fontannaz}
is based on collinear factorisation at NLO and the ZL prediction on a
$k_{T}$-factorisation approach~\cite{lipatov}. They equally investigated prompt
photon plus jet data are found to be reasonably well described by both
predictions, except at the highest $x_{\gamma}^{obs}$, where direct
photoproduction is   enhanced.

\section{Precision Physics with Jets}

\begin{figure}[h]
\centerline{\includegraphics[width=0.9\columnwidth]{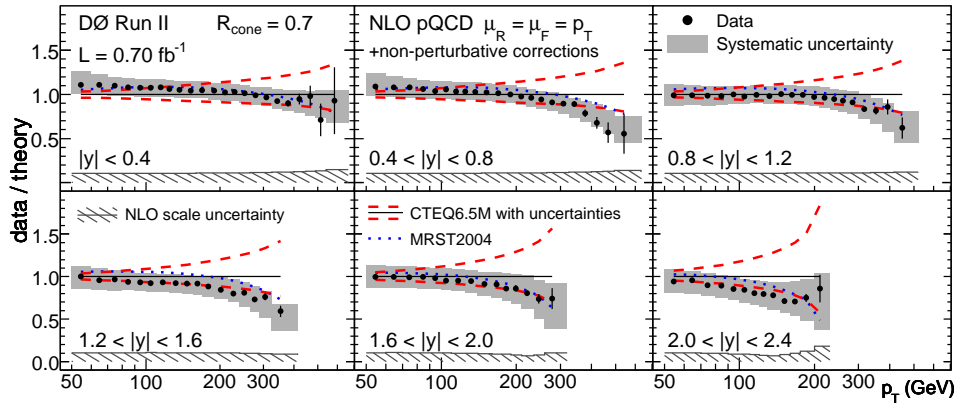}}
\caption{Inclusive jet cross section as measured by DZero divided by theory as a function of $p_T$ 
for 6 $|y|$ bins. Measurement uncertainties are shown by the shaded bands and PDF 
uncertainties by the dashed lines. The theoretical prediction using the MRST2004 PDFs is 
shown by the dotted line. The hashed area at the bottom of the plots shows the
uncertainty when varying the renormalisation and factorisation scales between $p_T/2$ and 
$2p_T$.  Figure taken
from~\protect\cite{mikko}.}\label{fig:mikko}
\end{figure}

Inclusive jet cross sections at $p\bar{p}$ colliders are a sensitive probe of parton 
distribution functions and of new physics. The DZero experiment has made the most
precise measurement of the inclusive jet cross section to 
date~\cite{mikko}. Figure~\ref{fig:mikko} 
shows the ratio of the measured to predicted inclusive jet cross sections as a 
function of jet $p_T$ in six rapidity bins. The theoretical prediction comes from a NLO 
QCD calculation using the CTEQ6.5 PDFs. The MRST2004 PDF predictions are also
plotted. As is demonstrated in this figure, the 
experimental uncertainties are now smaller than the scale and PDF uncertainties. The
dominant experimental uncertainty comes from the jet energy scale, and after seven years 
of work, the DZero collaboration has reduced this uncertainty to about 1.2\% for high $p_T$
central jets.

%\end{wrapfigure}

Studies are ongoing at the LHC in preparation for data taking in late 2008. New physics can be seen
as one goes to higher $p_T$ in the inclusive jet spectrum.  The 
current limit on the contact
interaction scale  at the Tevatron  is 2.7 TeV. 
The CMS experiment at the LHC
will be able~\cite{schieferdecker}
 to improve this limit within the first 10~pb$^{-1}$ of collected data. 
Assuming a jet energy scale
uncertainty of 10\% the predicted sensitivity to interaction scale is $\Lambda \approx $ 3 TeV, 7 TeV and 10 TeV
for 10~pb$^{-1}$, 100~pb$^{-1}$ and 1~fb$^{-1}$ respectively.

\begin{wrapfigure}{r}{0.5\columnwidth}
\centerline{\includegraphics[totalheight=8cm,width=0.4\columnwidth]{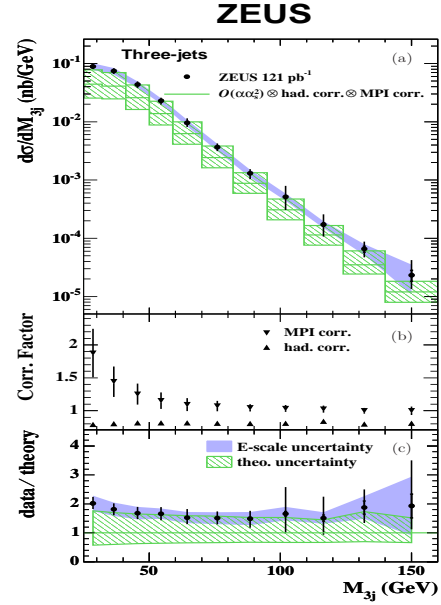}}
\caption{Photoproduction cross section of 3-jets as a function of the 3-jet invariant-mass at HERA. Figure taken from~\protect\cite{namsoo}.}\label{fig:namsoo}
\end{wrapfigure}
Recent measurements by the ZEUS collaboration of 2,3 and 4 jets in
photoproduction were presented~\cite{namsoo}. Compared to earlier 2-jet
results, the jet transverse energy requirements ($E_{T}^{jet1(2)} > 20 \,
(15)$~GeV), the range in pseudorapidity ($-1 < \eta^{jet} < 3$) and the
luminosity have been significantly increased in order to obtain a dataset
suitable for testing and constraining both the photon and the proton PDFs. The
sensitivity to different PDFs was shown by comparing the data to different
available parametrisations. For the 3 and 4-jets analysis the selection of the
jet phase space was different: $E_{T}^{jet} > 6$~GeV and $|\eta^{jet}| <
2.4$. In addition, the data were divided into a low and high mass sample with
$25 < M_{nj} < 50$~GeV and $M_{nj} > 50$~GeV in order to look for effects of
multi-parton interactions or underlying events and to allow for tests of
multi-jet description in event generators. In general both models HERWIG and
PYTHIA largely underestimate the rate of multi-jet events. At high $M_{nj}$
however, they provide a reasonable description of the shape of the data, while
low $M_{nj}$ and low $x_{\gamma}^{obs}$ are only described once multi-parton
interactions are introduced.  
A leading order QCD calculation for 3-jets, corrected for hadronisation
effects and effects of multi-parton interactions, 
is compared to data in Figure~\ref{fig:namsoo}, illustrating the need for a
NLO calculation.  

\begin{wrapfigure}{r}{0.5\columnwidth}
\centerline{\includegraphics[width=0.45\columnwidth]{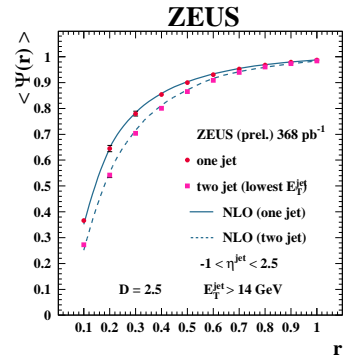}}
\caption{Integrated jet shapes for quark and gluon enriched jets in DIS at 
HERA. Figure taken from~\protect\cite{ron}.}\label{fig:ron}
\end{wrapfigure}
For jets with sufficiently high $E_{T}^{jet}$ one may expect the contribution
of fragmentation to the jet substructure to become small compared to the
contribution from parton radiation. Preliminary measurements by the ZEUS
collaboration~\cite{ron}
 of the mean integrated jet shape $\langle \Psi(r) \rangle$ 
for two
types of jets with $14 < E_{T}^{jet} < 17$~GeV are in good agreement with
corresponding NLO calculations as illustrated in Figure~\ref{fig:ron}. The
data are also well reproduced by the colour dipole model as implemented in
ARIADNE. The ``one-jet'' distribution is mainly due to quark-initiated jets,
selected by demanding events with 1-jet only. The ``two-jet'' distribution is
enriched with gluon-initiated jets by choosing the jet with the lowest
$E_{T}^{jet}$ in events which have only  two jets, which are close to each other
in $\eta-\phi$ space. One observes that the gluon enriched jets are broader as
expected. 

\begin{wrapfigure}{r}{0.5\columnwidth}
\epsfig{file= 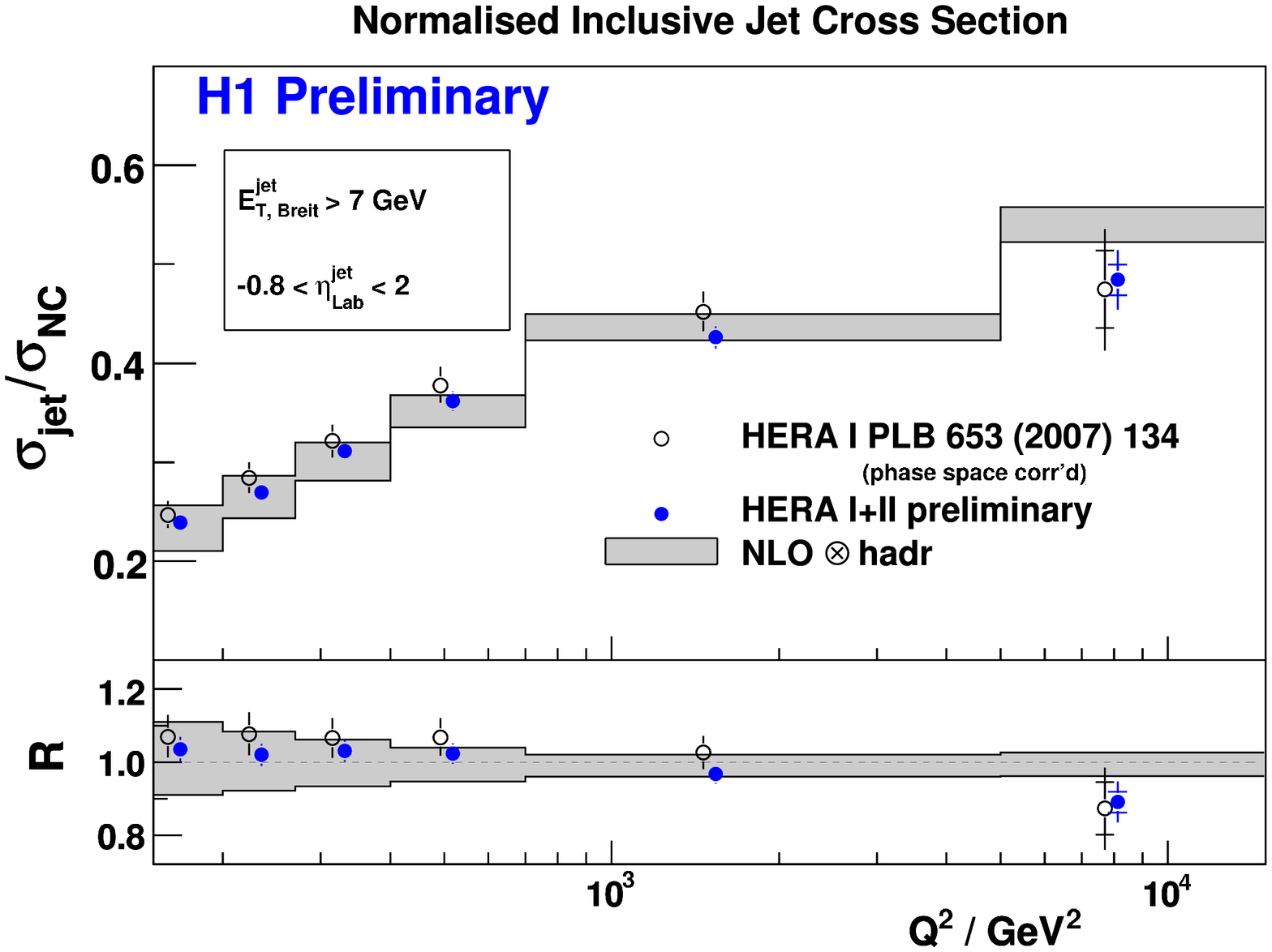,width=0.3\columnwidth,angle=-90,clip}
%\centerline{\includegraphics[width=0.3\columnwidth,angle=-90,clip]
%{gouzevitch1_mod}}
\caption{Normalised inclusive jet cross section and ratio of data over theory
at HERA. Figure taken from~\protect\cite{gouzevitch}.}\label{fig:gouzevitch1}
\end{wrapfigure}
Recent polarised and unpolarised inclusive jet cross sections were measured by the ZEUS collaboration in charged current (CC) $e^{\pm}p$ events using HERA-II data~\cite{theedt}. They are found to be in good agreement with Standard Model predictions. 
These jet cross sections and the fully inclusive CC cross sections can provide
constraints on the $u$ and $d$-PDFs at high x.  

Preliminary results on normalised inclusive, 2-jet and 3-jet cross sections by
the H1 collaboration were presented~\cite{gouzevitch}, using HERA-I and II
data. In photon virtuality they cover the range $5 < Q^{2} < 15000$~GeV$^{2}$
and in jet transverse energies $7 (5) < E_{T}^{jet} < 50 (80)$~GeV. The
normalised inclusive jet cross section for the high $Q^{2}$data, i.e. $Q^{2} >
150$~GeV$^{2}$, is shown in Figure~\ref{fig:gouzevitch1} together with
published HERA-I data and NLO QCD calculations~\cite{nagy}, illustrating good
agreement and experimental errors, which are clearly smaller than the
uncertainty of the NLO calculation. These high-$Q^{2}$ data have reached 
an experimental uncertainty of about $3$~\%, which is mainly due to the
uncertainty of the jet energy scale of about $1.5$~\%. 

\begin{wrapfigure}{r}{0.5\columnwidth}
\epsfig{file=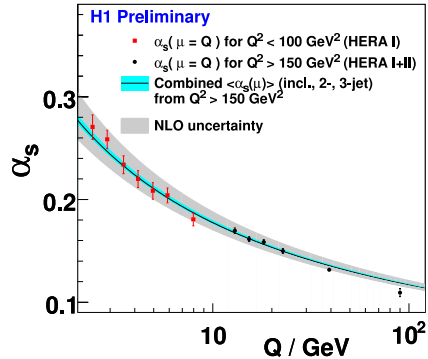,width=0.35\columnwidth,angle=-90,clip}
%\centerline{\includegraphics[width=0.35\columnwidth,angle=-90,clip]{gouzevitch2_mod}}
\caption{Running of $\alpha_{s}$ from jet cross sections at HERA. Figure
taken from~\protect\cite{gouzevitch}.}\label{fig:gouzevitch2}
\end{wrapfigure}
The value of the strong coupling %\begin{displaymath}
\begin{eqnarray*}
 \alpha_{s}(M_{Z}) = 0.1182 \;\pm\; 0.0008 \;\mathrm{(exp)} \;\\
 ^{+0.0041}_{-0.0031} \;\mathrm{(theo)} \;\pm\; 0.0018 \;\mathrm{(PDF)}
\end{eqnarray*}
%\end{displaymath} 
was extracted from a fit of the NLO calculation to the normalised inclusive, 2 and 3-jet cross sections, resulting in an impressively small experimental error of $0.7$~\%. The total uncertainty is dominated by the uncertainty of the NLO theory, indicating the need for a NNLO calculation. The running of the strong coupling is verified over two orders of magnitude in $Q$ as demonstrated in 
Figure~\ref{fig:gouzevitch2}.

\begin{wrapfigure}{r}{0.5\columnwidth}
\centerline{\includegraphics[width=0.45\columnwidth]{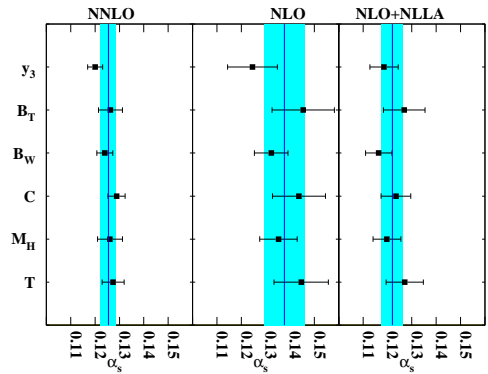}}
\caption{Strong coupling constant extracted from ALEPH 
data on event shapes. Figure taken
from~\protect\cite{stenzel}.}\label{fig:stenzel}
\end{wrapfigure}

Using the recently computed NNLO
 corrections to event shape 
variables \cite{eventshapes}, 
a new extraction of $\alpha_s$ from data on the standard set of 
six event shape variables, measured~\cite{ALEPH-qcdpaper} 
 by the ALEPH\ collaboration at LEP1 and LEP2 was performed.
One observes a clear improvement in the fit quality when going to
NNLO accuracy~\cite{stenzel}. 
Compared to NLO the value of $\alpha_s$ is lowered 
by about 10\%, but still higher than for NLO matched with 
the resummed next-to-leading 
logarithmic approximation (NLO+NLLA)~\cite{ALEPH-qcdpaper}.
As illustrated in Figure~\ref{fig:stenzel}, the scatter among the
 $\alpha_s$-values extracted from different shape variables is 
lowered considerably, and the theoretical uncertainty is decreased by 
a factor 2 (1.3) compared to NLO (NLO+NLLA). 
The combination of 
all shape variables at all energies yields 
\begin{displaymath}
    \alpha_s(M_Z) = 0.1240 \;\pm\; 0.0008\,\mathrm{(stat)}
     					 \;\pm\; 0.0010\,\mathrm{(exp)}
                                   \;\pm\; 0.0011\,\mathrm{(had)}
                                   \;\pm\; 0.0029\,\mathrm{(theo)} .
 \end{displaymath}

The fixed-order QCD description of event shape distributions is reliable 
only if the event shape variable is sufficiently far
away from its two-jet limit. In the approach to this limit, event shapes
display large infrared logarithms at all orders in perturbation theory, such
that the expansion in the strong coupling constant fails to converge.
Resummation of these logarithms yields a
description appropriate to the two-jet limit. To explain event shape
distributions over their full kinematical range, both descriptions need to be
matched onto each other. The matching of NLLA with NNLO was performed 
recently~\cite{luisoni}.  The most
striking observation is that the difference between NLLA+NNLO
and NNLO is largely restricted to the two-jet region, while
NLLA+NLO differ in normalisation throughout the full kinematical range.
This behaviour may serve as a first indication for the
numerical smallness of corrections beyond NNLO in the three-jet region.
Fits of $\alpha_s$ based on NNLO+NLLA are currently in progress.

\begin{wrapfigure}{r}{0.5\columnwidth}
\centerline{\includegraphics[width=0.35\columnwidth,angle=-90]{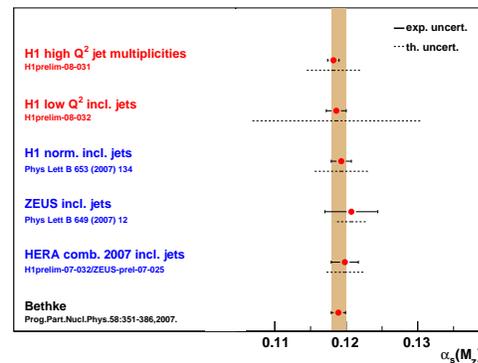}}
\caption{Summary of extractions of $\alpha_s(M_Z)$ at DIS 2008.}
\label{fig:alphas-dis08}
\end{wrapfigure}
A summary of the most precise $\alpha_s(M_Z)$ extractions 
from HERA data 
discussed at DIS 2008 is shown in 
Figure~\ref{fig:alphas-dis08}. It also includes the ``HERA comb.\ 2007
incl.\ jets'' result~\cite{kluge} 
of a first simultaneous fit to published HERA-I inclusive jet cross 
section data from H1 and ZEUS. 

\section{New Jet Algorithms}

The reconstruction of hadronic jets at colliders proceeds through the 
application of a jet algorithm, which clusters the detector-level 
information on the hadrons (hadronic tracks or calorimeter energy deposits) 
into a limited number of composite objects: the jets. One distinguishes 
two types of jet algorithms: cone-based  and cluster-based algorithms. 
The cone-based algorithms aim to maximise the hadronic energy inside a 
cone of fixed size, while the cluster-based algorithms perform a sequential 
recombination of parton pairs. For both classes of algorithms, many different 
realisations have been proposed in the past. Depending on the 
observable under consideration, one or the other  type of algorithm 
may be more appropriate: while cone-type algorithms perform generally better 
in the reconstruction of resonances, i.e.\ in searches for massive particles 
decaying into jets, cluster-type algorithms are more appropriate for 
precision studies. To compare 
experimental jet measurements with perturbative QCD, any algorithm 
of either type must 
fulfil infrared safety criteria.

Unfortunately, most of the cone-based algorithms used up to now 
at the Tevatron did not fulfil those criteria and display infrared 
sensitivity above a certain final-state jet multiplicity. 
The principal cause of this infrared sensitivity is the use of cone seeds 
to speed up the jet reconstruction. The seedless infrared-safe 
cone algorithm (SISCone)~\cite{soyez}  overcomes these problems and provides 
a cone-type algorithm for hadron collider physics. 

The sequential recombination used in clustering-based algorithms is very 
time-consuming, since the distance 
measure for each pair of (pseudo-)particles has to be evaluated in each 
iteration. The practical applicability of these algorithms was therefore 
severely restricted,
especially at hadron colliders, where typical events contain a 
very high multiplicity of hadronic objects. Using techniques from 
computational geometry, the clustering can be performed in a much more 
efficient way~\cite{fastkt}, such that cone-based and clustering 
algorithms display similar performance~\cite{rojo}.

By generalising the $k_T$-clustering algorithm to allow arbitrary powers 
in the transverse momentum weight, one can define a one-parameter family 
of clustering algorithms. The weighting with the inverse power of the 
transverse momentum defines the anti-$k_T$ algorithm~\cite{antikt}. A new
powerful tool to analyse the features of jets is their catchment 
area~\cite{jetarea}, which is obtained by including 
zero-energy ghost particles in the clustering. By inspecting the jet areas,
one observes that the   anti-$k_T$ algorithm yields perfect cones, which 
makes this algorithm an ideal replacement for iterative cone-type algorithms. 

\begin{wrapfigure}{r}{0.5\columnwidth}
\centerline{\includegraphics[width=0.45\columnwidth]{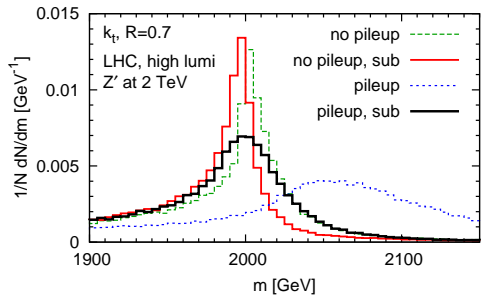}}
\caption{Effect of jet area based subtraction on reconstruction 
of heavy resonance at $M=2$~TeV decaying into jets. Figure taken
from~\protect\cite{jetarea}.}\label{fig:jetarea}
\end{wrapfigure}
The jet area may be turned into a powerful tool to disentangle the hard 
scattering process from any underlying event activity. On an event-by-event 
basis, this underlying activity, which appears to be  uniform in rapidity,
could be measured in areas outside any jet, and then subtracted from the 
 event prior to reconstruction. To illustrate the improvements which
can be obtained using this procedure, Figure~\ref{fig:jetarea} displays the 
effect of event pileup on the reconstruction of the new heavy $Z'$ 
gauge boson decaying into hadronic jets. While the mass peak is 
substantially shifted by the effect of pileup, it is observed at its 
true mass after jet area based subtraction.

\section{Conclusions}
The broad spectrum of new results on ``Hadronic final states and QCD'' 
presented in the working group, as well as the very lively discussions,
illustrate
the importance of this subject area in the current transition period between 
HERA and the LHC. Many recent experimental results are of direct relevance 
to preparations for LHC physics, and theoretical physics is responding to
a number of challenges in complexity and precision posed by the physics
programme of present and future colliders.

\section*{Acknowledgment}

TG would like to thank the Swiss National Science Foundation (SNF) for support 
under contract 200020-117602.

% ****************************************************************************
% BIBLIOGRAPHY AREA
% ****************************************************************************

\begin{footnotesize}
% IF YOU DO NOT USE BIBTEX, USE THE FOLLOWING SAMPLE SCHEME FOR THE REFERENCES
% ----------------------------------------------------------------------------

\end{footnotesize}

\begin{thebibliography}{99}

\bibitem{cooper}
  T.~Aaltonen {\it et al.}, [CDF collaboration],
 Phys.\ Rev.\ Lett.\  {\bf 100} (2008) 102001  
 and B.\ Cooper, these proceedings.
 
\bibitem{dobson}
E.\ Dobson, these proceedings.


\bibitem{kauer}
  T.~Binoth, M.~Ciccolini, N.~Kauer and M.~Kramer,
  %``Gluon-induced W-boson pair production at the LHC,''
  JHEP {\bf 0612} (2006) 046
  [hep-ph/0611170];\\
  %%CITATION = JHEPA,0612,046;%%
 T.~Binoth, N.~Kauer and P.~Mertsch, these proceedings
  %``Gluon-induced QCD corrections to pp --> ZZ --> l anti-l l' anti-l',''
  [arXiv:0807.0024].
  %%CITATION = ARXIV:0807.0024;%%

\bibitem{chachamis}
 G.~Chachamis, M.~Czakon and D.~Eiras,
  %``W Pair Production at the LHC - I. Two-loop Corrections in the High Energy
  %Limit,''
  arXiv:0802.4028; 
  %%CITATION = ARXIV:0802.4028;%%
  %``W Pair Production at the LHC - II. One-loop Squared Corrections in the High
  %Energy Limit,''
  arXiv:0806.3043
  %%CITATION = ARXIV:0806.3043;%%
and G.~Chachamis, these proceedings.

\bibitem{bartels}
 J.~Bartels, M.~Salvadore and G.P.~Vacca,
  %``Inclusive 1-jet Production Cross Section at Small x in QCD: Multiple
  %Interactions,''
  JHEP {\bf 0806} (2008) 032
  [arXiv:0802.2702].
  %%CITATION = JHEPA,0806,032;%% 


\bibitem{jimmy}
 I.~Borozan and M.H.~Seymour,
  %``An eikonal model for multiparticle production in hadron hadron
  %interactions,''
  JHEP {\bf 0209} (2002) 015
  [hep-ph/0207283].
  %%CITATION = JHEPA,0209,015;%%

\bibitem{baehr1}
 M.~B\"ahr, J.M.~Butterworth and M.H.~Seymour,
  %``The Underlying Event and the Total Cross Section from Tevatron to the
  %LHC,''
  arXiv:0806.2949.
  %%CITATION = ARXIV:0806.2949;%%


\bibitem{metha}
A.\ Metha, these proceedings.


\bibitem{bechtel}
F.\ Bechtel, these proceedings.

\bibitem{herwig}
  M.~B\"ahr {\it et al.},
  %``Herwig++ Physics and Manual,''
  arXiv:0803.0883.
  %%CITATION = ARXIV:0803.0883;%%


\bibitem{baehr2}
M.~B\"ahr, S.~Gieseke and M.H.~Seymour,
  %``Simulation of multiple partonic interactions in Herwig++,''
  JHEP {\bf 0807} (2008) 076  [arXiv:0803.3633]
  %%CITATION = ARXIV:0803.3633;%%
and M.~B\"ahr, these proceedings.

\bibitem{sherpa}
 T.~Gleisberg, S.~H\"oche, 
F.~Krauss, A.~Schalicke, S.~Schumann and J.C.~Winter,
  %``SHERPA 1.alpha, a proof-of-concept version,''
  JHEP {\bf 0402} (2004) 056
  [hep-ph/0311263] and F.\ Krauss, these proceedings.
  %%CITATION = JHEPA,0402,056;%%

\bibitem{sherpa1}
 A.~Schalicke and F.~Krauss,
  %``Implementing the ME+PS merging algorithm,''
  JHEP {\bf 0507} (2005) 018
  [hep-ph/0503281].
  %%CITATION = JHEPA,0507,018;%%

\bibitem{sherpa2}
  S.~Schumann and F.~Krauss,
  %``A parton shower algorithm based on Catani-Seymour dipole factorisation,''
  JHEP {\bf 0803} (2008) 038
  [arXiv:0709.1027];\\
  %%CITATION = JHEPA,0803,038;%%
 J.C.~Winter and F.~Krauss,
  %``Initial-state showering based on colour dipoles connected to incoming
  %parton lines,''
 JHEP {\bf 0807} (2008) 040    [arXiv:0712.3913].
  %%CITATION = ARXIV:0712.3913;%%

\bibitem{sherpa3}
 T.~Gleisberg and F.~Krauss,
  %``Automating dipole subtraction for QCD NLO calculations,''
  Eur.\ Phys.\ J.\  C {\bf 53} (2008) 501
  [arXiv:0709.2881].
  %%CITATION = EPHJA,C53,501;%%

\bibitem{sherpa4}
 C.~Duhr, S.~H\"oche and F.~Maltoni,
  %``Color-dressed recursive relations for multi-parton amplitudes,''
  JHEP {\bf 0608} (2006) 062
  [hep-ph/0607057] and S.\ H\"oche, these proceedings.
  %%CITATION = JHEPA,0608,062;%%



\bibitem{mcnlo}
  S.~Frixione and B.~R.~Webber,
  %``Matching NLO QCD computations and parton shower simulations,''
  JHEP {\bf 0206} (2002) 029
  [hep-ph/0204244];\\
  %%CITATION = JHEPA,0206,029;%%
  S.~Frixione, P.~Nason and C.~Oleari,
  %``Matching NLO QCD computations with Parton Shower simulations: the POWHEG
  %method,''
  JHEP {\bf 0711} (2007) 070
  [arXiv:0709.2092];\\
  %%CITATION = JHEPA,0711,070;%%
  Z.~Nagy and D.~E.~Soper,
  %``Matching parton showers to NLO computations,''
  JHEP {\bf 0510} (2005) 024
  [hep-ph/0503053].
  %%CITATION = JHEPA,0510,024;%%

\bibitem{sherstnev}
  A.~Sherstnev and R.S.~Thorne,
  %``Parton Distributions for LO Generators,''
  Eur.\ Phys.\ J.\  C {\bf 55} (2008) 553
  [arXiv:0711.2473] and A.~Sherstnev, these proceedings.
  %%CITATION = EPHJA,C55,553;%%

\bibitem{khein}
 S.~Chekanov {\it et al.}  [ZEUS Collaboration],
  %``Forward-jet production in deep inelastic ep scattering at HERA,''
  Eur.\ Phys.\ J.\  C {\bf 52} (2007) 515
  [arXiv:0707.3093] and L.\ Khein, these proceedings.
  %%CITATION = EPHJA,C52,515;%%



\bibitem{h1-fwdjets}
  A.~Aktas {\it et al.}  [H1 Collaboration],
  %``Forward jet production in deep inelastic scattering at HERA,''
  Eur.\ Phys.\ J.\  C {\bf 46} (2006) 27
  [hep-ex/0508055].
  %%CITATION = EPHJA,C46,27;%%


\bibitem{turnau}
J.\ Turnau, these proceedings.

\bibitem{hautmann}
F.\ Hautmann, these proceedings.

\bibitem{royon}
 O.~Kepka, C.~Royon, C.~Marquet and R.B.~Peschanski,
  %``Next-to-leading BFKL phenomenology of forward-jet cross sections at HERA,''
  Eur.\ Phys.\ J.\  C {\bf 55} (2008) 259
  [hep-ph/0612261] and C.~Royon, these proceedings. 
  %%CITATION = EPHJA,C55,259;%%
\bibitem{kutak}
K.\ Kutak, these proceedings.

\bibitem{zotov}
  S.P.~Baranov, A.V.~Lipatov and N.P.~Zotov,
  %``Prompt photon hadroproduction at high energies in off-shell gluon-gluon
  %fusion,''
  Phys.\ Rev.\  D {\bf 77} (2008) 074024
  [arXiv:0708.3560] and N.P.~Zotov, these proceedings.
  %%CITATION = PHRVA,D77,074024;%%

\bibitem{saleev}
V.A.\ Saleev, these proceedings.


\bibitem{hoeche}
S.~H\"oche, F.~Krauss and T.~Teubner,
  %``Multijet events in the k_T-factorisation scheme,''
  arXiv:0705.4577 and S.\ H\"oche, these proceedings.
  %%CITATION = ARXIV:0705.4577;%%

\bibitem{white}
 J.R.~Andersen and C.D.~White,
  %``A New Framework for Multijet Predictions and its application to Higgs Boson
  %production at the LHC,''
  arXiv:0802.2858 and C.D.\ White, these proceedings.
  %%CITATION = ARXIV:0802.2858;%%


\bibitem{kidonakis}
 N.~Kidonakis, A.\ Sabio Vera and P.~Stephens,
  %``Resummations in QCD hard-scattering at large and small x,''
  arXiv:0802.4240 and N.~Kidonakis, these proceedings.
  %%CITATION = ARXIV:0802.4240;%%


\bibitem{karshon}  S.~Chekanov {\it et al.}  [ZEUS Collaboration],
arXiv:0806.0807, to be published in Phys.\ Rev.\ Lett.\ 
and U.~Karshon, these proceedings.

\bibitem{TASSO_L3} M.~Althoff {\it et al.}
 [TASSO Collaboration], Phys.\ Lett.\ B 
{\bf 121}
 (1983) 216;\\ 
M.~Acciarri {\it et al.} [L3 Collaboration],
 Phys.\ Lett.\ B {\bf 501} (2001) 173.

\bibitem{devita}
R.\ De Vita, these proceedings.

\bibitem{gilfoyle}
G.P.\ Gilfoyle, these proceedings.

\bibitem{caines}
 A.~Adare {\it et al.} [Phoenix Collaboration],
 Phys.\ Rev.\ C {\bf 78} (2008) 014901  
[arXiv:0801.4545] and H.~Caines, these proceedings.
  


\bibitem{ZEUSchp}S.~Chekanov {\it et al.}
 [ZEUS Collaboration], JHEP {\bf 0806} (2008) 061 
and T.~Tymieniecka, these proceedings. 

\bibitem{H1chp} F.D.~Aaron {\it et al.} [H1 Collaboration], 
Phys.\ Lett.\ B {\bf 654} 
(2007) 148 
and D.~Traynor, these proceedings.

\bibitem{falkiewicz} A.~Falkiewicz, these proceedings. 

\bibitem{B-Jmodel} A.~Bia{\l}as and M.~Je\.{z}abek, Phys.\ Lett.\ B {\bf 590}
 (2004) 233.

\bibitem{tymieniecka} T.~Tymieniecka, these proceedings.

\bibitem{gapienko} G.~Gapienko, these proceedings.


\bibitem{moch}
 S.~Moch and A.~Vogt,
  %``On Third-Order Timelike Splitting Functions and Top-Mediated Higgs Decay
  %into Hadrons,''
  Phys.\ Lett.\  B {\bf 659} (2008) 290
  [arXiv:0709.3899] and S.\ Moch, these proceedings.
  %%CITATION = PHLTA,B659,290;%%

\bibitem{kniehl}
S.~Albino, B.A.~Kniehl and G.~Kramer,
  %``AKK Update: Improvements from New Theoretical Input and Experimental
  %Data,''
Nucl.\ Phys.\ B {\bf 803} (2008) 42
  [arXiv:0803.2768] and B.~Kniehl, these proceedings.
  %%CITATION = ARXIV:0803.2768;%%

 \bibitem{campanelli}
V.M.~Abazov {\it et al.} [DZero Collaboration],
  arXiv:0804.1107 and M.\ Campanelli, these proceedings. 
\bibitem{jetphox}
  P.~Aurenche, M.~Fontannaz, J.P.~Guillet, E.~Pilon and M.~Werlen,
  %``A new critical study of photon production in hadronic collisions,''
  Phys.\ Rev.\  D {\bf 73} (2006) 094007
  [hep-ph/0602133].
  %%CITATION = PHRVA,D73,094007;%%

  
\bibitem{nowak}
K.\ Nowak, these proceedings.

\bibitem{fontannaz}
M.~Fontannaz, J.~P.~Guillet and G.~Heinrich,
  %``Isolated prompt photon photoproduction at NLO,''
  Eur.\ Phys.\ J.\  C {\bf 21} (2001) 303
  [hep-ph/0105121]; \\
  %%CITATION = EPHJA,C21,303;%%
M.~Fontannaz and G.~Heinrich,
  %``Isolated photon + jet photoproduction as a tool to constrain the gluon
  %distribution in the proton and the photon,''
  Eur.\ Phys.\ J.\  C {\bf 34} (2004) 191
  [hep-ph/0312009].
  %%CITATION = EPHJA,C34,191;%%  

\bibitem{lipatov}
A.V.~Lipatov and N.~P.~Zotov,
  %``Prompt photon photoproduction at HERA in the k(T)-factorization
  %approach,''
  Phys.\ Rev.\  D {\bf 72} (2005) 054002
  [hep-ph/0506044].
  %%CITATION = PHRVA,D72,054002;%%  

  
  
\bibitem{mikko}
V.M.~Abazov {\it et al.} [DZero Collaboration],
  arXiv:0802.2400 and M.\ Voutilainen, these proceedings. 

\bibitem{schieferdecker}
P.\ Schieferdecker, these proceedings.

\bibitem{namsoo}
T.\ Namsoo, these proceedings.

\bibitem{ron}
E.\ Ron, these proceedings.

\bibitem{theedt}
T.\ Theedt, these proceedings.

\bibitem{gouzevitch}
M.\ Gouzevitch, these proceedings.

\bibitem{nagy}
Z.~Nagy and Z.~Trocsanyi,
  %``Multi-jet cross sections in deep inelastic scattering at next-to-leading
  %order,''
  Phys.\ Rev.\ Lett.\  {\bf 87} (2001) 082001
  [hep-ph/0104315].
  %%CITATION = PRLTA,87,082001;%%


  



\bibitem{eventshapes}
 A.~Gehrmann-De Ridder, T.~Gehrmann, E.W.N.~Glover and G.~Heinrich,
  %``NNLO corrections to event shapes in $e^+e^-$ annihilation,''
  JHEP {\bf 0712} (2007) 094
  [arXiv:0711.4711].
  %%CITATION = JHEPA,0712,094;%%

\bibitem{ALEPH-qcdpaper}
A.~Heister {\it et al.}  [ALEPH Collaboration],
  %``Studies of QCD at e+ e- centre-of-mass energies between 91-GeV and
  %209-GeV,''
  Eur.\ Phys.\ J.\  C {\bf 35} (2004) 457.
  %%CITATION = EPHJA,C35,457;%%


\bibitem{stenzel}
 G.~Dissertori, A.~Gehrmann-De Ridder, T.~Gehrmann, E.W.N.~Glover, G.~Heinrich 
and H.~Stenzel,
  %``First determination of the strong coupling constant using NNLO predictions
  %for hadronic event shapes in e^+e^- annihilations,''
  JHEP {\bf 0802} (2008) 040
  [arXiv:0712.0327] and H.\ Stenzel, these proceedings. 
  %%CITATION = JHEPA,0802,040;%%

\bibitem{luisoni}
 T.~Gehrmann, G.~Luisoni and H.~Stenzel,
  %``Matching NLLA+NNLO for event shape distributions,''
  Phys.\ Lett.\  B {\bf 664} (2008) 265
  [arXiv:0803.0695] and G.~Luisoni, these proceedings.
  %%CITATION = PHLTA,B664,265;%%

\bibitem{kluge}
T.\ Kluge, these proceedings.  


\bibitem{soyez}
 G.P.~Salam and G.~Soyez,
  %``A practical Seedless Infrared-Safe Cone jet algorithm,''
  JHEP {\bf 0705} (2007) 086
  [arXiv:0704.0292].
  %%CITATION = JHEPA,0705,086;%%

\bibitem{fastkt}
 M.~Cacciari and G.P.~Salam,
  %``Dispelling the N**3 myth for the k(t) jet-finder,''
  Phys.\ Lett.\  B {\bf 641} (2006) 57
  [hep-ph/0512210].
  %%CITATION = PHLTA,B641,57;%%

\bibitem{rojo}
J.\ Rojo, these proceedings [arXiv:0806.3958].
  %%CITATION = ARXIV:0806.3958;%%

\bibitem{antikt}
 M.~Cacciari, G.P.~Salam and G.~Soyez,
  %``The anti-k_t jet clustering algorithm,''
  JHEP {\bf 0804} (2008) 063
  [arXiv:0802.1189] and G.\ Soyez, these proceedings.
  %%CITATION = JHEPA,0804,063;%%

\bibitem{jetarea}
 M.~Cacciari and G.P.~Salam,
  %``Pileup subtraction using jet areas,''
  Phys.\ Lett.\  B {\bf 659} (2008) 119
  [arXiv:0707.1378];\\
  %%CITATION = PHLTA,B659,119;%%
 M.~Cacciari, G.P.~Salam and G.~Soyez,
  %``The Catchment Area of Jets,''
  JHEP {\bf 0804} (2008) 005
  [arXiv:0802.1188] and M.\ Cacciari, these proceedings.
  %%CITATION = JHEPA,0804,005;%%





\end{thebibliography}
\end{document}